\documentclass[aps,pra,twocolumn,floatfix,superscriptaddress]{revtex4}

\usepackage{mathrsfs}
\usepackage{graphicx}
\usepackage[english]{babel}
\usepackage{amsmath}
\usepackage{amssymb}

\usepackage{relsize}
\usepackage{dsfont}
\usepackage{bm}

\newcommand{\me}{\mathrm{e}}
\newcommand{\mi}{\mathrm{i}}

\newcommand{\dif}{\mathrm{d}}

\allowdisplaybreaks
\begin{document}

\title{Ubiquity of zeros of Loschmidt amplitude for mixed states in different physical processes and their implications}

\author{Xu-Yang Hou}
\affiliation{Department of Physics, Southeast University, Jiulonghu Campus, Nanjing 211189, China}
\author{Qu-Cheng Gao}
\affiliation{Department of Physics, Southeast University, Jiulonghu Campus, Nanjing 211189, China}
\author{Hao Guo}
\email{guohao.ph@seu.edu.cn}
\affiliation{Department of Physics, Southeast University, Jiulonghu Campus, Nanjing 211189, China}
\author{Yan He}
\affiliation{College of physics, Sichuan University, Chengdu, Sichuan 610064, China}
\author{Tong Liu}
\email{t6tong@njupt.edu.cn}
\affiliation{Department of Applied Physics, Nanjing University of Posts and Telecommunications, Nanjing 210003, China}

\author{Chih-Chun Chien}
\email{cchien5@ucmerced.edu}
\affiliation{Department of physics, University of California, Merced, CA 95343, USA}

\begin{abstract}
The Loschmidt amplitude of the purified states of mixed-state density matrices is shown to have zeros when the system undergoes a quasistatic, quench, or Uhlmann process. While the Loschmidt-amplitude zero of a quench process corresponds to a dynamical quantum phase transition (DQPT) accompanied by the diverging dynamical free energy, the Loschmidt-amplitude zero of the Uhlmann process corresponds to a topological phase transition (TQPT) accompanied by a jump of the Uhlmann phase. Although the density matrix remains intact in a quasistatic process, the Loschmidt amplitude can have zeros not associated with a phase transition. We present examples of two-level and three-level systems exhibiting finite- or infinite- temperature DQPTs and finite-temperature TQPTs associated with the Loschmidt-amplitude zeros. Moreover, the  dynamical phase or geometrical phase of mixed states can be extracted from the Loschmidt amplitude. Those phases may become quantized or exhibit discontinuity at the Loschmidt-amplitude zeros. A spinor representation of the purified states of a general two-level system is presented to offer more insights into the change of purification in different processes. The quasistatic process, for example, is shown to cause a rotation of the spinor.
\end{abstract}

\maketitle

\section{Introduction}
In recent years, two classes of phase transitions, the topological quantum phase transitions (TQPTs)~\cite{WXG95,KaneRMP,ZhangSCRMP,ChiuRMP}, where nonanalytic behavior exists in the topology of the system, and the dynamical quantum phase transitions (DQPTs)~\cite{Zvyagin16,DQPTreview18}, where nonanalytic behavior exists in the dynamics of the system, have attracted broad research interest. Although the literature mainly focuses on the ground-state transitions, generalizations of the two types of transitions to mixed states in or out of equilibrium are challenges that need to be addressed in order to provide a more complete picture of the physics behind those transitions.

A TQPT of the ground state usually refers to a change of a topological invariant associated with the Hamiltonian mapping or the band structure~\cite{KaneRMP,ZhangSCRMP,ChiuRMP}. Finding a proper topological invariant to characterize the TQPT of mixed states may be more complicated. There have been attempts to use various geometric phases to achieve the goal. For example, the Uhlmann phase~\cite{Uhlmann86,Uhlmann89,Uhlmann91,Uhlmann96} is thought of as a generalization of the Berry phase~\cite{Berry84} to mixed states. Another example is the interferometric geometric phase~\cite{GPMQS1} inspired by the Mach-Zehnder interferometry. Finite-temperature TQPTs have been studied in some quantum systems in Refs.~\cite{ViyuelaPRL14,2DMat15} by monitoring the dependence of the Uhlmann phase on temperature, where a qualitative change of the Uhlmann phase is claimed to signify a finite-temperature TQPT. However, complications may arise when dynamical processes are involved. Although a quantum system described by a pure state can simultaneously acquire the geometrical Berry phase and the dynamical phase during a single adiabatic process, Ref.~\cite{ourPRB20} shows the Uhlmann phase and the dynamical phase can not be generated concurrently during a single dynamical process due to the incompatibility between the Uhlmann process based on the concept of parallel transport and the dynamical process governed by the Hamiltonian.

On the other hand, the DQPTs reveal the nonanalytic behavior in real-time dynamics of quantum systems. There have been extensive studies~\cite{DQPT13,DQPT14,DQPT15,DQPTreview18} and progresses in both the theories \cite{DQPTB2,DQPTB3,DQPTB4a0,DQPTB4a,DQPTB4b,DQPTB4c,DQPTB4d0,DQPTB4d,DQPTB4d2,DQPTB4e1,DQPTB4e,DQPTB5,JafariSciRep19} and experiments \cite{DQPTB41,DQPTB4}. An important tool in the study of DQPTs is the Loschmidt amplitude. The DQPTs occur at the zeros of the Loschmidt amplitude, which are called the Fisher zeros when the real time is complexified on the complex plane. The Loschmidt amplitude
provides an analogue of the partition function of the thermodynamic phase transitions of quantum systems. While the majority of investigations of the DQPTs focuses on pure quantum states, there have been generalizations to mixed quantum states~\cite{DQPTM16,DTQPT18,DQPTM17} and open systems~\cite{Bandyopadhyay18,DQPTM18}.
There may be several ways to generalize the concept of the Loschmidt amplitude to mixed states~\cite{DQPTreview18,Sedlmayr18}.
Here we follow the generalization in which the Loschmidt amplitude is defined as the overlap between the purified states of the density matrices because this approach also applies to the TQPTs, as will be shown later. We consider systems with short-range interactions and mention that DQPTs of some systems with long-range interactions have been studied in Refs.~\cite{Homrighausen17,LangPRB18,LangPRL18}. A common feature of the zero-temperature TQPTs and DQPTs may be seen as follows. The $T=0$ TQPTs are usually characterized by topological invariants~\cite{ChiuRMP}. Take the 1D Su-Schrieffer-Heeger (SSH) model for example, the the winding number is associated with the 1D Berry phase, also known as the Zak phase~\cite{Asboth2016}. Since the Uhlmann phase approaches the Berry phase as $T\rightarrow 0$, a discontinuous geometric phase signifies the TQPT in the SSH model. On the other hand, the zero-temperature DQPTs correspond to the vanishing wavefunction overlap and causes non-analytic behavior in the dynamical phase~\cite{Zvyagin16,DQPTreview18}.

Here we will address the question regarding whether the TQPTs and DQPTs of mixed states share something in common by showing the transition points of the mixed-state TQPTs and DQPTs are both zeros of the corresponding Loschmidt amplitude of the purified states. The association of the Loschmidt-amplitude zeros with TQPTs has not been broadly recognized in the literature. Our approach not only explains the change of the Uhlmann phase across a TQPT~\cite{ViyuelaPRL14,2DMat15}, but also provides a physical picture of the underlying mechanism by the vanishing Loschmidt amplitude at the TQPT. Moreover, we will illustrate that the Loschmidt amplitude can have zeros in other dynamical processes by analyzing the quasistatic process, which is a typical process in thermodynamics~\cite{Schroeder_book,ourPRB20}. Interestingly, the density matrix remains the same in a quasistatic process, so there is no phase transition. Nevertheless, the purified states may vary and lead to vanishing Loschmidt amplitude even when the system  follows a quasistatic process. A unified picture of the zeros of the Loschmidt amplitude in different processes provides a deeper understanding of the purification of density matrix and connects the seemingly different DQPTs and TQPTs. Moreover, we will unveil the mathematical structure behind the purified states of a generic two-level system by showing a spinor representation. The quasistatic process is equivalent to a rotation of the spinor representing the purified state, which explains why the Loschmidt amplitude may vanish even though the density matrix remains the same.

There have been experimental realizations of some DQPTs of the ground states~\cite{DQPTreview18} and excited states~\cite{TianPRL20}, and there have been experimental implications of the TQPTs indicated by the jump of the Uhlmann phase~\cite{npj18}. Nevertheless, there remain many unsolved questions about the Uhlmann process and dynamical process of mixed states, especially when it comes to the internal structure of the density matrix and its geometric properties. The unified view of the Loschmidt-amplitude zero presented here may offer clues of a unified description of the topology and dynamics of mixed states.

The rest of the paper is organized as follows. In Sec.~\ref{Sec1}, we give an overview of purification of a density matrix and discuss the resulting amplitude and purified state, laying the foundation for the rest of the paper. In Sec.~\ref{Sec2}, we introduce three physical processes that a mixed state can experience, including the quasistatic, quench, and Uhlmann processes. In Sec.~\ref{Sec3}, we give explicit examples to analyze the zeros of the Loschmidt amplitudes in those three processes. In Sec.~\ref{RP}, we discuss a spinor representation of purification of two-level systems. Sec.~\ref{Sec5} concludes our work.

\section{Purification of density Matrix }\label{Sec1}
\subsection{Purification}
The discussions of the TQPTs and DQPTs of mixed states are based on the concept of purification of density matrices.
In general, a density matrix of a mixed quantum state can be decomposed as
\begin{align}\label{rhoW}
\rho=WW^\dagger,
\end{align}
where $W$ is called the amplitude of $\rho$. 
The amplitude of a density matrix is not uniquely determined because the amplitude can be conversely expressed as
$W=\sqrt{\rho}U$, where $U$ is a unitary matrix. If $W$ is full rank, this polar decomposition of $W$ is unique, and the corresponding density matrix is called ``faithful''~\cite{Uhlmann86}. In the rest of the paper we will focus on full-rank density matrices unless specified otherwise.
Thus, the amplitude $W$ plays the role of a wavefunction and $U$ is the generalization of the $U(1)$ phase factor of the wavefunction. The amplitudes
form a Hilbert space $H_W$, where a scalar product, called the Hilbert-Schmidt product, is defined as $(W_1,W_2):=\textrm{Tr}(W^\dagger_1W_2)$ ~\cite{Uhlmann89}.
This is the overlap between two amplitudes, which may also be referred to as the quantum fidelity between two purifications.
In the Hilbert space $\mathcal{H}$ spanned by the eigenvectors of $\rho$,
the amplitude is expressed as $W=\sum_i\sqrt{\lambda_i}|i\rangle\langle i|U$,
where $\lambda_i$ is the $i$-th eigenvalue of $\rho$.

The purification can also be cast into the form of a pure state by introducing an ancilla. The procedure introduces an isomorphism between the spaces $H_W$ and $\mathcal{H}\otimes\mathcal{H}$ as follows.
\begin{align}\label{w2}
W=\sum_i\sqrt{\lambda_i}|i\rangle\langle i|U\leftrightarrow|W\rangle=\sum_i\sqrt{\lambda_i}|i\rangle\otimes U^T|i\rangle,
\end{align}
where $U^T$ is the transpose of $U$ taken with respect to the eigenbasis of $\rho$ and acts on the aforementioned ancilla, or the second Hilbert space. The purified state can be constructed by doubling the degrees of freedom via
an auxiliary system with an identical Hilbert space.
Although the two terminologies, purified state and amplitude, are usually used interchangeably in most of the literature, we will use $W$ exclusively for the amplitude and $|W\rangle$ for the purified state from the purification. It can be shown that the inner product between two purified states gives rise to the Hilbert-Schmidt product $\langle W_1|W_2\rangle=\textrm{Tr}(W^\dagger_1W_2)=(W_1,W_2)$~ \cite{HubnerPLA93}.
The density matrix $\rho$ can be recovered by tracing out the auxiliary degrees of freedom of the enlarged space $\mathcal{H}\otimes\mathcal{H}$. Explicitly,
\begin{align}\label{dme1}
\rho=\textrm{Tr}_2(|W\rangle\langle W|),
\end{align}
where $\textrm{Tr}_2$ is the partial trace taken over the second Hilbert space.

It can be shown that the expectation value of an arbitrary observable $\mathcal{O}$ with respect to the purified state $|W\rangle$ is the same as its statistical average in the mixed state described by $\rho=WW^\dagger$. Hence,
\begin{align}\label{Oav}
\bar{\mathcal{O}}\equiv \langle \mathcal{O}\rangle =\text{Tr}(\rho\mathcal{O})=\langle W|\mathcal{O}|W\rangle.
\end{align}
To verify the relation, we note that $\mathcal{O}$ acts on the first Hilbert space of $\mathcal{H}\otimes\mathcal{H}$. The trace in Eq.~(\ref{Oav}) is in fact $\text{Tr}_1$, i.e., the partial trace taken over the first Hilbert space. By using Eq.~(\ref{dme1}), we have
$\text{Tr}_1\left[\mathcal{O}\textrm{Tr}_2(|W\rangle\langle W|)\right]=\langle W|\mathcal{O}|W\rangle$.
If $\rho$ describes an ensemble in thermal equilibrium with well-defined temperature, Eq.~(\ref{Oav}) shows that its purification is a thermal-vacuum state~\cite{QP_book,Vitiello_book}.

\subsection{Parallelity and orthogonality of pure and mixed states}
In quantum information theory, the overlap between two pure quantum states is called the quantum fidelity, which is a measure of the difference between the two states~\cite{WatrousBook}. Following Ref.~\cite{Uhlmann86}, two pure states $|\psi_1\rangle$ and $|\psi_2\rangle$ are said to be parallel to each other if
\begin{align}\label{WPT3}
\langle \psi_1|\psi_2\rangle=\langle \psi_2|\psi_1\rangle>0,
\end{align}
i.e. the associated fidelity is a positive real number. Under this condition, the Fubini-Study distance between the two states is minimized, indicating that the ``difference'' between them is minimal.
Physically, two parallel states have maximal quantum fidelity because the fidelity provides a quantitative measure of
how close two states of a quantum system are to each other.
On the other hand, two pure states are said to be orthogonal to each other if
\begin{align}\label{WPT30}
\langle \psi_1|\psi_2\rangle=0.
\end{align}
In other words, the two states have minimal similarity since any one of them contains no component of the other. In this situation, the violation of the parallelity is maximal.

Both the parallelity and orthogonality relations between pure states can be generalized to mixed states by using $\langle W_1|W_2\rangle=\textrm{Tr}(W^\dagger_1W_2)$. Two purified states are orthogonal to each other if $\langle W_1|W_2\rangle=0$. However, this condition cannot give a deterministic relation between $\rho_1$ and $\rho_2$.
The generalization of the parallelity condition is more involved.
It seems a direct generalization to the condition (\ref{WPT3}) is
\begin{align}\label{WPT4c}
\langle W_1|W_2\rangle=\langle W_2|W_1\rangle>0.
\end{align}
Instead, Uhlmann imposed a stronger condition: Two amplitudes satisfy the parallel condition $W_1\parallel W_2$ if
 \begin{align}\label{WPT4a}
W^\dagger_2W_1=W^\dagger_1W_2>0
\end{align}
where $W^\dagger_1W_2>0$ means the eigenvalues of $W^\dagger_1W_2$ are all positive real numbers.
This Uhlmann condition immediately implies Eq.~(\ref{WPT4c}), and the spectrum of the operator (\ref{WPT4a}) is an invariant of the ordered pair $\rho_1,\rho_2$~\cite{Uhlmann86}.
Similar to the pure-state case, the Hilbert-Schmidt distance between $W_1$ and $W_2$ is minimized if they are parallel to each other.

\section{Loschmidt amplitude and physical Processes}\label{Sec2}
Here we analyze selected physical processes that a mixed state may experience, including two dynamical processes and the Uhlmann process. The two types of dynamical processes analyzed here are the quench process and the quasistatic process. We show that the Loschmidt amplitude can exhibit zeros in all the three processes, but the implications and interpretations of the zeros are different.

\subsection{Dynamical processes}

\subsubsection{Quasistaic process}\label{qdp}
Among the physical processes discussed here, the quasistatic process, where any intermediate state is an equilibrium state, is probably the simplest one. The quasistatic process is commonly assumed in thermodynamics textbooks~\cite{Schroeder_book}. For simplicity, we assume that a system evolves dynamically according to a time-independent Hamiltonian $H$ during a quasistatic process. The time evolution of the density matrix is described by the Heisenberg equation
\begin{align}\label{Heq}
\mi\hbar\dot{\rho}=[H,\rho],
\end{align}
which is formally solved by
\begin{align}\label{sHeqs}
\rho(t)=\me^{-\frac{\mi}{\hbar}Ht}\rho(0)\me^{\frac{\mi}{\hbar}Ht}.
\end{align}
The quasistatic condition requires $[\rho,H]=0$, implying the density matrix remains unchanged. However, later we will show that nontrivial results can emerge even in this simple situation.
Comparing Eq.~\eqref{sHeqs} to Eq.~(\ref{rhoW}), we have
\begin{align}\label{Wt0}
W(t)W^\dagger(t)&=\me^{-\frac{\mi}{\hbar}Ht}W(0)W^\dagger(0)\me^{\frac{\mi}{\hbar}Ht}\notag\\&=\me^{-\frac{\mi}{\hbar}Ht}W(0)\left[\me^{-\frac{\mi}{\hbar}Ht}W(0)\right]^\dagger.
\end{align}
Thus,
\begin{align}\label{Wt1}
W(t)=\me^{-\frac{\mi}{\hbar}Ht}W(0),
\end{align}
which solves the Schr$\ddot{\text{o}}$dinger's equation of the amplitude,
\begin{align}\label{WSe}
\mi\hbar \dot{W}(t)=HW(t).
\end{align}
Hence, $W(t)$ is indeed the amplitude of $\rho(t)$. As time elapses, the corresponding purified state evolves according to 
\begin{align}\label{Wt}
|W(t)\rangle=\me^{-\frac{\mi}{\hbar}Ht}\otimes 1|W(0)\rangle.
\end{align}
Here $\me^{-\mi Ht}$ acts on the first Hilbert space and the identity operator $1$ acts on the second (auxiliary) Hilbert space.
Note that $W(0)W^\dagger(0)=\rho(0)=\rho(t)=W(t)W^\dagger(t)$, i.e. $W(0)$ and $W(t)$ are two different amplitudes of the same density matrix in the quasistatic process.

To extract information between the initial and final amplitudes in a physical process, we employ the Loschmidt amplitude~\cite{DQPTreview18}, defined as
\begin{align}\label{lamqs}
\mathcal{G}_\rho(t)=\langle W(0)|W(t)\rangle=\text{Tr}(\rho(0)\me^{-\frac{\mi}{\hbar}Ht}).
\end{align}
The Loschmidt amplitude reveals the overlap between two purified states, which can be viewed as a generalization of the quantum fidelity to mixed quantum states.
In general, $\mathcal{G}_\rho$ is a complex number, and its argument is the dynamical phase of the corresponding process~\cite{ourPRB20}:
\begin{align}\label{DPqp}
\theta_D(t)=\arg\left[\mathcal{G}_\rho(t)\right].
\end{align}
An interesting question is whether $\mathcal{G}_\rho(t)$ of a quasistatic process can possess zeros and what is the physical meaning and implication? We remark that if the zeros exist in a quasistatic process, the same density matrix can have two (or more) different purified states orthogonal to each other. This implies the purification allows the introduction of additional information than the density matrix itself.

\subsubsection{Quench process}
Different from the quasistatic process that keeps the system in equilibrium at each instance, a sudden change of the Hamiltonian is exerted in the beginning of a quench process.
We assume the system is initially prepared
in a mixed state described by the density matrix $\rho(0)$. The initial state may be a nonequilibrium mixed state or an equilibrium ensemble with well-defined temperature.
At time $t=0^+$, the Hamiltonian is suddenly switched to a new Hamiltonian $H_\text{f}$ and the system evolves according to the new Hamiltonian. In general, $[\rho(t), H_\text{f}]\neq 0$, making the quench process different from the quasistatic process.

Similar to the analysis of the quasistatic process, we assume $H_\text{f}$ is time-independent. By replacing $H$ by $H_\text{f}$ in Eq.~(\ref{sHeqs}), the time evolution of the density matrix follows the expression
\begin{align}\label{sHe}
\rho(t)=\me^{-\frac{\mi}{\hbar}H_\text{f}t}\rho(0)\me^{\frac{\mi}{\hbar}H_\text{f}t}.
\end{align}
Following a similar derivation of Eq.~(\ref{Wt0}), the amplitude is shown to evolve according to
\begin{align}\label{Wtqs1}
W(t)=\me^{-\frac{\mi}{\hbar}H_\text{f}t}W(0), \text{ or equivalently },  \mi\hbar \dot{W} =H_\text{f} W.
\end{align}
The Loschmidt amplitude has been defined in Eq.~(\ref{lamqs}) and can be applied to the quench process as well~\cite{DQPTreview18}.
It has been argued that the Loschmidt amplitude can be thought of as the real-time generalization of the partition function of an equilibrium ensemble. Similar to the quasistatic process, the argument of the Loschmidt amplitude is the dynamical phase of the corresponding process, given by $\theta_D(t)=\arg\left[\mathcal{G}_\rho(t)\right]$.

The zeros of the Loschmidt amplitude correspond to the DQPTs by the analogy between the Loschmidt amplitude and the partition function~\cite{DQPTreview18}. At the critical point of a DQPT, the purified state after the quench shares the minimal similarity to the initial purified state. Since a quench process is generically a nonequilibrium process, the conventional definition of the thermodynamic free energy does not apply here. Nevertheless, the dynamical free-energy density of a dynamic process may be defined as
\begin{equation}\label{eq:fenergy}
f(t)=-\lim_{L\rightarrow \infty}\frac{1}{L}\ln \mathcal{L}_\rho(t),
\end{equation}
where $\mathcal{L}_\rho(t)=|\mathcal{G}_\rho(t)|^2$ is the Loschmidt echo~\cite{DQPTreview18}, and $L$ is the overall degrees of freedom of the system. From this point of view, the DQPTs at the zeros of $\mathcal{G}_\rho(t)$ correspond to the non-analytic points of the dynamical free-energy density. We emphasize that both $f(t)$ and $\theta_D$ exhibit non-analytic behavior at a DQPT. Explicit examples will be presented later.

\subsection{Uhlmann Process}\label{QP}
Our discussions of the TQPTs will be focused on selected topological systems going through the Uhlmann processes.
Here we summarize the Uhlmann process and its associated Uhlmann phase with minimal reference to the complex mathematical language of fiber bundles. Instead, we use the previously introduced concept of the parallelity between quantum states~\cite{ourPRB20}. An Uhlmann process is a cyclic process, during which the amplitude of the density matrix is parallel-transported. Here a cyclic process means $\rho(0)=\rho(\tau)$, where $\tau$ marks the end of a cycle of the Uhlmann process. However, the parallel transport may cause $W(\tau)$ to deviate from $W(0)$. From the difference between the two amplitudes, the Uhlmann phase can be extracted and represent a generalization of the Aharanov-Anandan phase for mixed states~\cite{GPbook}. We remark that the Uhlmann process is a process incompatible with the dynamical process governed by the Hamiltonian~\cite{ourPRB20}. If we parametrize the Uhlmann process by a variable, it should not be identified as the time. Therefore, we will use $s$ as the parameter of an Uhlmann process, and $\tau$ should not be misunderstood as a period of time.

To clarify the physical meaning of the parallel transport, we set $W_1\equiv W(0)$ and $W_2\equiv W(0^+)$ in Eq.~(\ref{WPT4a}). The differential form of the parallel-transport condition is then~\cite{ourPRB20}
 \begin{align}\label{WPT}
\dot{W}^\dagger W=W^\dagger \dot{W},
\end{align}
where $\dot{W}\equiv \frac{\dif W(s)}{\dif s}$. The expression can be cast into the form of an ``anti-Hermitian Schrodinger equation''
\begin{align}\label{Ha}
\mi\hbar\dot{W}=\tilde{H}W,
\end{align}
where $\tilde{H}=\mi\hbar\dot{W}W^{-1}$ is an anti-Hermitian matrix. Note that $\tilde{H}$ is the generator of the ``evolution'' during the Uhlmann processes. Moreover, note that
\begin{align}\label{Ha3}
\dot{W}^\dagger W=\frac{\mi}{\hbar}W^\dagger \tilde{H}^\dagger W=-\frac{\mi}{\hbar}W^\dagger \tilde{H} W=W^\dagger \dot{W}.
\end{align}
Thus, it follows from $\rho=WW^\dagger$ that
\begin{align}\label{Urhod}
\dot{\rho}=\dot{W}W^\dagger+W\dot{W}^\dagger=-\frac{\mi}{\hbar}\{\tilde{H},\rho\}.
\end{align}
The non-hermiticity of $\tilde{H}$ indicates that the normalization $\text{Tr}\rho(s)=1$ may be violated when $0<s<\tau$ and there may be other complications~\cite{Serg14}.
Therefore, the Uhlmann process cannot be achieved in a closed quantum system.
This is in stark contrast to the dynamical processes discussed previously.

During an Uhlmann process, the amplitude of the density matrix is parallel-transported according to Eq.~(\ref{WPT}).
An interesting question is if $W_2$ is obtained by a parallel transport of $W_1$ along a path through a \textit{finite} distance, is $W_2$ still parallel to $W_1$? Note that Eq.~(\ref{WPT4a}) defines a binary relation. However, it is not an equivalence relation since it possesses reflexivity and symmetry but lacks transivity. In other words, $W_1\parallel W_2$ and $W_2\parallel W_3$ $ \nRightarrow$
$W_1\parallel W_3$. This is because the space, where the amplitudes live, is the Uhlmann bundle, which is generally a curved space. The existence of the Uhlmann curvature leads to the failure of the transivity.

Here is a question regarding the Uhlmann process: Is it possible that the parallelity between the amplitudes can be completely lost during an Uhlmann process? If it is possible, the Loschmidt amplitude $\mathcal{G}^U_\rho(T,\tau)\equiv\langle W(0)|W(\tau)\rangle$ will have zeros.
For an Uhlmann process, one can infer from Refs.~\cite{Uhlmann86,Uhlmann89,Uhlmann91} that
 \begin{align}\label{GrU}
\mathcal{G}^U_\rho(T,\tau)=\text{Tr}(\rho(0)\mathcal{P}\me^{-\oint_\tau A_U}),
\end{align}
where $\mathcal{P}$ is the path-ordering operator, and $A_U$ is the Uhlmann connection given by
\begin{align}\label{GrAU}
A_U=-\sum_{ij}|i\rangle\frac{\langle i|[\mathrm{d}\sqrt{\rho},\sqrt{\rho}]|j\rangle}{\lambda_i+\lambda_j}\langle j|.
\end{align}
The argument of $\mathcal{G}^U_\rho(T,\tau)$ is the Uhlmann phase:
\begin{align}\label{GPU}
\theta_U=\arg\text{Tr}(\rho(0)\mathcal{P}\me^{-\oint_\tau A_U}).
\end{align}
Note that the incompatibility between the dynamical process and Uhlmann process actually resolves the puzzle regarding how one can differentiate the geometrical phase from the dynamical phase when the phase is extracted from the argument of the Loschmidt amplitude after going through a single physical process.
For pure states, a simultaneous generation of the dynamical phase and Berry phase are allowed~\cite{ourPRB20}.
For mixed states, the incompatibility between the Uhlmann and dynamical processes ensures that the argument of the Loschmidt amplitude can only contribute to one type of the phases: Either the dynamical phase or the Uhlmann phase will be accumulated, depending on the process that the system undergoes.

Finally, we want to point out that the Uhlmann phase carries the geometrical information of the associated system since the initial and final amplitudes are connected by an Uhlmann holonomy element~\cite{Uhlmann86,UP2D15}. Therefore, $\mathcal{G}^U_\rho$ and $\theta_U$ in general do not depend on the ``length'' of $\tau$ as long as the cycle is completed. We also caution that the Uhlmann bundle is trivial~\cite{DiehlPRB15}, limiting the type of topological invariants of the bundle. The Uhlmann phase may be considered as the geometric phase acquired by the purified state when the system follows the Uhlmann process, during which the purified state is parallel-transported~\cite{ViyuelaPRL14,Asorey19}. By mapping out the density matrix and reconstructing the overlap~\cite{npj18}, the Uhlmann phase may be measured experimentally.

\section{Examples of Loschmidt-amplitude zeros of mixed states}\label{Sec3}
After showing the universal mechanism of vanishing Loschmidt amplitude behind the DQPTs and TQPTs of mixed states, we will use generic two-level and three-level systems to demonstrate the ubiquity of the zeros of the Loschmidt amplitude of mixed states after the systems undergo a quasistatic, quench, or Uhlmann process. The examples will also answer the questions posed in previous sections.

\subsection{Two-level system}
\subsubsection{Quasistatic process}
We first consider a generic two-level system experiencing a quasistatic dynamical process.
The Hamiltonian has the form $H=\mathbf{R}\cdot \vec{\sigma}$, where $\mathbf{R}$ is a real-valued vector and $\vec{\sigma}=(\sigma_x,\sigma_y,\sigma_z)^T$ are the Pauli matrices. During a quasistatic process, the system remains in equilibrium at each instance and thus has a well-defined temperature.
Let $\beta=(k_B T)^{-1}$ and we will choose $k_B=1$. We define $R=|\mathbf{R}|$, $\hat{\mathbf{R}}=\mathbf{R}/R$, and $\omega=\frac{R}{\hbar}$. The time-evolution operator and initial density matrix are respectively given by
 \begin{align}
\me^{-\frac{\mi}{\hbar}Ht}&=\cos(\omega t)-\mi\sin(\omega t)\hat{\mathbf{R}}\cdot\vec{\sigma},\label{H2l}\\
\rho(0)&=\frac{1}{2}\left(1-\tanh(\beta R)\hat{\mathbf{R}}\cdot\vec{\sigma}\right).\label{rho2l}
\end{align}
By substituting those results into Eq.~(\ref{lamqs}), the Loschmidt amplitude becomes
\begin{align}\label{lamqs1}
\mathcal{G}_\rho(T,t)=\cos(\omega t)+\mi\sin(\omega t)\tanh(\beta R).
\end{align}
The expression only becomes zero if the temperature goes to infinity, or $\beta\rightarrow 0$. Moreover, the zeros occur when
\begin{align}\label{t0}
t^*=\frac{\left(n+\frac{1}{2}\right)\pi}{\omega}=\frac{\hbar\left(n+\frac{1}{2}\right)\pi}{R},
\end{align}
where $n$ is an integer.
We caution that the zeros of $\mathcal{G}_\rho$ from a quasistatic process do not correspond to a phase transition. This is because the system, being in equilibrium in each instance, has a well-behaved thermodynamic free energy $F=-\frac{1}{\beta}\ln Z$ that has no nonanalytic behavior at those zeros of $\mathcal{G}_\rho$. Nevertheless, the zeros from a quasistatic process do reveal some interesting aspects of the representation of the purified state of a mixed state, which will be discussed later.

The dynamical phase may look ill-defined when $\mathcal{G}_\rho$ vanishes. However, one must be careful  when evaluating $\theta_D$ because the limit should only be taken in the last step. A careful calculation shows that
\begin{align}\label{DPH2iT}
\lim_{T\rightarrow \infty}\theta_D\left(T,t=\frac{\left(n+\frac{1}{2}\right)\pi}{\omega}\right)=(-1)^{n}\frac{\pi}{2}.
\end{align}
Moreover, the dynamical phase can be shown to always take discrete values at infinite temperature. The time $t^*$ of the zeros are called the ``resonant points'' in Ref.~\cite{ourPRB20}, where the dynamic phase at infinite temperature exhibits quantized jumps. When $t\neq t^*$, or the "ordinary points" in Ref.~\cite{ourPRB20}, the dynamical phase takes the following two constant values at infinite temperature:
\begin{widetext}
\begin{align}\label{DPH2iT2}
\lim_{T\rightarrow \infty}\theta_D\left(T,t\neq \frac{\left(n+\frac{1}{2}\right)\pi}{\omega}\right)=\left\{\begin{array}{cc} 0 & \text{if } \omega t\in \left(2n\pi-\frac{\pi}{2},2n\pi+\frac{\pi}{2}\right),\\
\pi  & \text{if }  \omega t\in \Big[(2n-1)\pi,2n\pi-\frac{\pi}{2}\Big)\cup\Big(2n\pi+\frac{\pi}{2},(2n+1)\pi\Big].
\end{array}\right.
\end{align}
\end{widetext}

The analysis above gives a positive answer to the question in Sec.\ref{qdp} on whether the Loschmidt amplitude can have zeros in a quasistatic process. However, there is no phase transition associated with the zeros.  We note that the time evolution of a quasistatic process is a unitary transformation of the amplitude of the density matrix, but the density matrix is unchanged. Thus, the initial purification can lose its characteristics in a quasistatic process while the succeeding purification becomes maximally different from the initial one at $t^*$ when the Loschmidt amplitude vanishes.

Our analysis of the Loschmidt-amplitude zeros of quasistatic processes suggests that purification may carry more information of a mixed state than its density matrix. Interestingly, it is known that different mixed states may be represented by the same density matrix~\cite{MQM}. In Sec.~\ref{RP}, we will give a detailed study of the representation of purification for two-level systems. The purified states can be thought of as spinors, and the quasistatic process is equivalent to a rotation acting on the spinors.

\subsubsection{Quench process}
The quench process has been extensively studied because of the DQPTs~\cite{DQPTreview18}, including two-level systems at finite temperatures~\cite{DQPTM17,DTQPT18}. Here we briefly discuss a two-level system through a quench process. The initial density matrix is assumed to have the form
\begin{align}\label{dofmin}
\rho(0)=\frac{1}{2}(1+\mathbf{R}_0\cdot \vec{\sigma}).
\end{align}
In general, the expression describes a nonequilibrium mixed state, so the temperature may not be well defined.
At $t=0^+$, the Hamiltonian is changed to $H_\text{f}=\mathbf{R}\cdot \vec{\sigma}$, and $[\rho(0),H_\text{f}]\neq 0$ (i.e., $\mathbf{R}_0\nparallel \mathbf{R}$). The final density matrix is usually different from the initial density matrix, marking an important difference from the quasistatic process.
Following the same convention as before and using Eq.~(\ref{lamqs}), we get
\begin{align}\label{H2LA}
\mathcal{G}_\rho(t)
=\cos(\omega t)-\mi\sin(\omega t)\mathbf{R}_0\cdot\hat{\mathbf{R}}.
\end{align}
When $\mathbf{R}_0\cdot\hat{\mathbf{R}}=0$, the Loschmidt amplitude possesses zeros at $t^*=\frac{\left(n+\frac{1}{2}\right)\pi}{\omega}$. It has been shown that DQPTs occur at the zeros of the Loschmidt amplitude~\cite{DQPTreview18}.
A special case is $\mathbf{R}_0=\mathbf{0}$, which corresponds to $\rho(0)=\frac{1}{2}1_{2\times 2}$. This can be recognized as the mixed state at infinite temperature.

If the initial mixed state is an equilibrium state, it has a well-defined temperature even though the temperature may no longer be well defined in the quench process. We assume the eigenvalues of the initial Hamiltonian are $\pm E$. The initial density matrix is then given by
\begin{align}\label{dofmi}
\rho(0)
=\frac{1}{Z(0)}\left[\cosh(\beta E)-\sinh(\beta E)\sigma_z\right].
\end{align}
Taking the eigenvectors of the initial Hamiltonian as the basis, we assume that the quenched Hamiltonian is $H_\text{f}=\mathbf{R}\cdot \vec{\sigma}$. The Loschmidt amplitude is then given by
 \begin{align}\label{GG3a}
&\mathcal{G}_\rho(T,t)=\frac{1}{Z(0)}\text{Tr}\big[\begin{pmatrix}\me^{-\beta E} & 0\\ 0 & \me^{\beta E}\end{pmatrix}\me^{-\mi\hat{\mathbf{R}}\cdot \vec{\sigma}\omega t}\big]\notag\\
&=\frac{2}{Z(0)}\left[\cosh(\beta E)\cos(\omega t)+\mi \frac{R_z}{R}\sinh(\beta E)\sin(\omega t)\right].
\end{align}
In the zero-temperature limit, the system reduces to the pure-state case, which have been studied thoroughly and summarized in Ref.~\cite{DQPTreview18}.

While a conventional phase transition occurs when the thermodynamic free energy exhibits nonanalytic behavior, a DQPT occurs when the Loschmidt amplitude vanishes~\cite{DQPTreview18}. One may wonder if there is a DQPT at finite temperature.
An analysis of Eq.~\eqref{GG3a} shows that the Loschmidt-amplitude zeros of a two-level system in a quench process can occur at $t^*=\dfrac{\left(n+\frac{1}{2}\right)\pi}{\omega}$ but only when $\beta=0$. Therefore, a two-level system can only have DQPTs at infinite temperature but not finite temperature, consistent with the result of Ref.~\cite{DTQPT18}. The reason a zero of the Loschmidt amplitude corresponds to a DQPT is because the dynamical free energy density, defined by
\begin{equation}\label{eq:dyfe}
f(t)=-\lim_{L\rightarrow \infty}\ln|\mathcal{G}_\rho(T\rightarrow\infty,t)|^2
\end{equation}
is singular as the Loschmidt amplitude vanishes. The dynamical free energy of a dynamical system is an analogue of the thermodynamic free energy of an equilibrium system~\cite{DQPTreview18}. Moreover, the dynamical phase at $t^*$ is $\theta_D=\text{sgn}\left(\frac{R_z}{R}\right)(-1)^n\frac{\pi}{2}$, showing a jump across a DQPT.

\subsubsection{Uhlmann process}
For the Uhlmann process, we consider a fermionic two-band system in 1D since some results are known~\cite{ViyuelaPRL14}. The Hamiltonian is expressed in terms of the Nambu representation $\Psi_k=(a_k, b_k)^T$, where $a_k$ and $b_k$ denote the fermionic
operators of two species, and
$k$ is the 1D crystalline momentum living
in a Brillouin
zone with the geometry of $S^1$.
The Hamiltonian has a
quadratic form $H=\sum_k\Psi^\dagger_kH_k\Psi_k$, where
\begin{align}\label{Hk2}
H_k=\frac{\Delta_k}{2}\hat{\mathbf{n}}_k\cdot\vec{\sigma}
\end{align}
with $\Delta_k$ determining the size of the energy gap.
The density matrix is obtained by assuming the system is in equilibrium, so $\rho=\prod_k\rho_k$ with
\begin{align}\label{rhok}
\rho_k=\frac{1}{2}\left(1-\tanh(\frac{\beta\Delta_k}{2})\hat{\mathbf{n}}_k\cdot\vec{\sigma}\right).
\end{align}
The 1D momentum $k$ itself can serve as the parameter $s$ for constructing the cyclic Uhlmann process.

There are two types of TQPTs involved here, one from the topology of the Hamiltonian mapping of the two-level system, and the other from the topology of the Uhlmann process of the two-level system.
We first discuss the TQPT associated with the Hamiltonian mapping. After $\rho_k$ is parallel-transported as $k$ traverse the Brillouin zone, the system acquires an Uhlmann phase. During the Uhlmann process, $\hat{\mathbf{n}}_k$ also varies. If the movement of $\hat{\mathbf{n}}_k$ is restricted on a plane, its tip lies on a circle, defining a mapping from $S^1\rightarrow S^1$. Thus, we can write $\hat{\mathbf{n}}_k$ as $\hat{\mathbf{n}}(k)$, which explicitly defines the map. The winding number from the mapping is given by
\begin{align}\label{wne}
\omega_{1}=\frac{1}{2 \pi} \oint\left(\frac{\partial_{k} n_{k}^{i}}{n_{k}^{j}}\right) \dif k.
\end{align}
Here $n_{k}^{i,j}$ are two nonzero components of $\mathbf{n}_k$.
In general, the winding number $\omega_1$ may take any integer values, depending on the explicit models. Here we consider two simple possibilities $\omega_1=1$ and $\omega_1=0$. A TQPT of the ground state occurs when $\omega_1$ changes its value. At finite temperatures, the Hamiltonian mapping does not take into account of the distribution function, and it is no longer an indicator of a TQPT.

On the other hand, there is a topology from the Uhlmann phase of a two-level system in a Uhlmann process. The Loschmidt amplitude $\mathcal{G}^U_\rho$ of the general two-band system~\eqref{Hk2} in an Uhlmann process can be inferred from  Ref.~\cite{ViyuelaPRL14}. Explicitly, we have 
\begin{align}\label{GU}
\mathcal{G}^U_\rho(T)=\cos \left(\pi \omega_{1}\right) \cos \left[\oint\left(\frac{\partial_{k} n_{k}^{i}}{2 n_{k}^{j}}\right) \operatorname{sech}\left(\frac{\Delta_{k}}{2k_B T}\right) \dif k\right].
\end{align}
A derivation is given in the Appendix.
Moreover, $\mathcal{G}^U_\rho(T)$ is real-valued for the two-level system, so the Uhlmann phase $\theta_U$ can only take the quantized values $0$ or $\pi$. Given  $\theta_U=\arg[\text{Tr}(W^\dagger(0)W(\tau))]$, the Uhlmann phase corresponds to the relative phase between the initial and final amplitudes. Since the Uhlmann process is a cyclic process, the parameter $s$ traverses a loop in the parameter space and the Uhlmann phase ends up being $0$ or $\pi$ after one cycle. Thus, the Uhlmann phase indicates whether the topology of the change of the purification is similar to a cylinder (with $\theta_U=0$ representing the identity element) or a Mobius strip (with $\theta_U=\pi$ representing a twist).
Therefore, a TQPT of mixed states occurs when the Uhlmann phase changes its value, which is possible when $\mathcal{G}^U_\rho(T^*)=0$ at some critical temperature $T^*$ because $\mathcal{G}^U_\rho(T)$ is a continuous function of $T$. Therefore, the zero of the Loschmidt amplitude of the Uhlmann process represents the boundary between the two different topological regimes, and the definition applies to finite-temperature TQPTs. If $\mathcal{G}^U_\rho=0$, $\theta_U$ becomes ill-defined. Thus, $\theta_U$ jumps from one value to another when crossing the critical temperature $T^*$, serving as an indicator of the finite-temperature TQPT. In Ref.~\cite{UP2D15}, the Uhlmann phase is said to be like a topological kink. Interestingly, the Matsubara formalism of finite temperature theory introduces the inverse temperature by a Wick rotation of the time \cite{MahanBook}. Here we also have the correspondence $\beta\hbar\sim \mi t$. Hence, the TQPTs may be recognized as the zeros of the Loschmidt amplitude with imaginary time while the DQPTs as the zeros of the Loschmidt amplitude with real time.

We notice that in the 1D two-level system, the zero-temperature TQPT characterized by $\omega_1$ affects the finite-temperature TQPT characterized by the zeros of the Loschmidt amplitude. The boundary of the finite-temperature TQPT is at the zeros of Eq.~\eqref{GU}.
If $T\rightarrow 0$, then $\operatorname{sech}\left(\frac{\Delta_{k}}{2k_B T}\right)\rightarrow 0$, and the argument of the second cosine function on the right-hand-side of Eq.~(\ref{GU}) is 0. If $T\rightarrow \infty$, then $\operatorname{sech}\left(\frac{\Delta_{k}}{2k_B T}\right)\rightarrow 1$, and the argument of the second cosine function becomes
\begin{align}
\lim_{T\rightarrow \infty}\oint\left(\frac{\partial_{k} n_{k}^{i}}{2 n_{k}^{j}}\right) \operatorname{sech}\left(\frac{\Delta_{k}}{2k_B T}\right) \dif k= \pi\omega_1.
\end{align}
Thus, the answer depends on $\omega_1$. When $\omega_1=1$, the range of the continuous function $\mathlarger{\oint}\left(\frac{\partial_{k} n_{k}^{i}}{2 n_{k}^{j}}\right) \operatorname{sech}\left(\frac{\Delta_{k}}{2k_B T}\right) \dif k$ is $[0,\pi)$ since $T\in [0,\infty)$. As a consequence, there must be at least an intermediate temperature $T^*$ such that
\begin{align}\label{T*}\oint\left(\frac{\partial_{k} n_{k}^{i}}{2 n_{k}^{j}}\right) \operatorname{sech}\left(\frac{\Delta_{k}}{2k_B T^*}\right) \dif k=\frac{\pi}{2}.\end{align}
At $T^{\ast}$ (or $\beta^*$), $\mathcal{G}^U_\rho(T^*)=0$, which  signifies a finite-temperature TQPT. The Uhlmann phase $\theta_U$ changes from $\pi$ (a topologically nontrivial phase) to $0$ (a topologically trivial phase) if $T$ crosses $T^{\ast }$.
On the other hand, when $\omega_1=0$, the map $\hat{\mathbf{n}}(k)$ is topologically trivial and the above argument fails. In this regime, there is no guarantee of the existence of any Loschmidt-amplitude zero at finite temperature.

Since the Uhlmann process is cyclic, $\rho(0)=\rho(\tau)$ but in general $\rho(0)\neq \rho(s)$ if $0<s<\tau$.
Here we emphasize again that the parameter $s$ can be chosen as $k$. Thus, $\tau$ is the ``size'' of the 1D Brillouin
zone, which is usually $2\pi$ in the suitable unit.
At the finite-temperature TQPT transition point $T^*$, the zero of the Loschmidt amplitude thus implies
 \begin{align}
\text{Tr}(W^\dagger(0)W(2\pi))=\langle W(0)|W(2\pi)\rangle=0.
\end{align}
Here $W(0)$ and $W(2\pi)$ are two amplitudes of the same density matrix. The situation is thus similar to the quasistatic process. The result also answers the question raised in Sec.~\ref{QP} because the final purification of a density matrix can maximally lose its similarity to the initial purification by showing orthogonality between the initial and final purified states, even though the purified state is parallel-transported in the Uhlmann process.

\begin{figure}[th]
\centering
\includegraphics[width=3.2in]{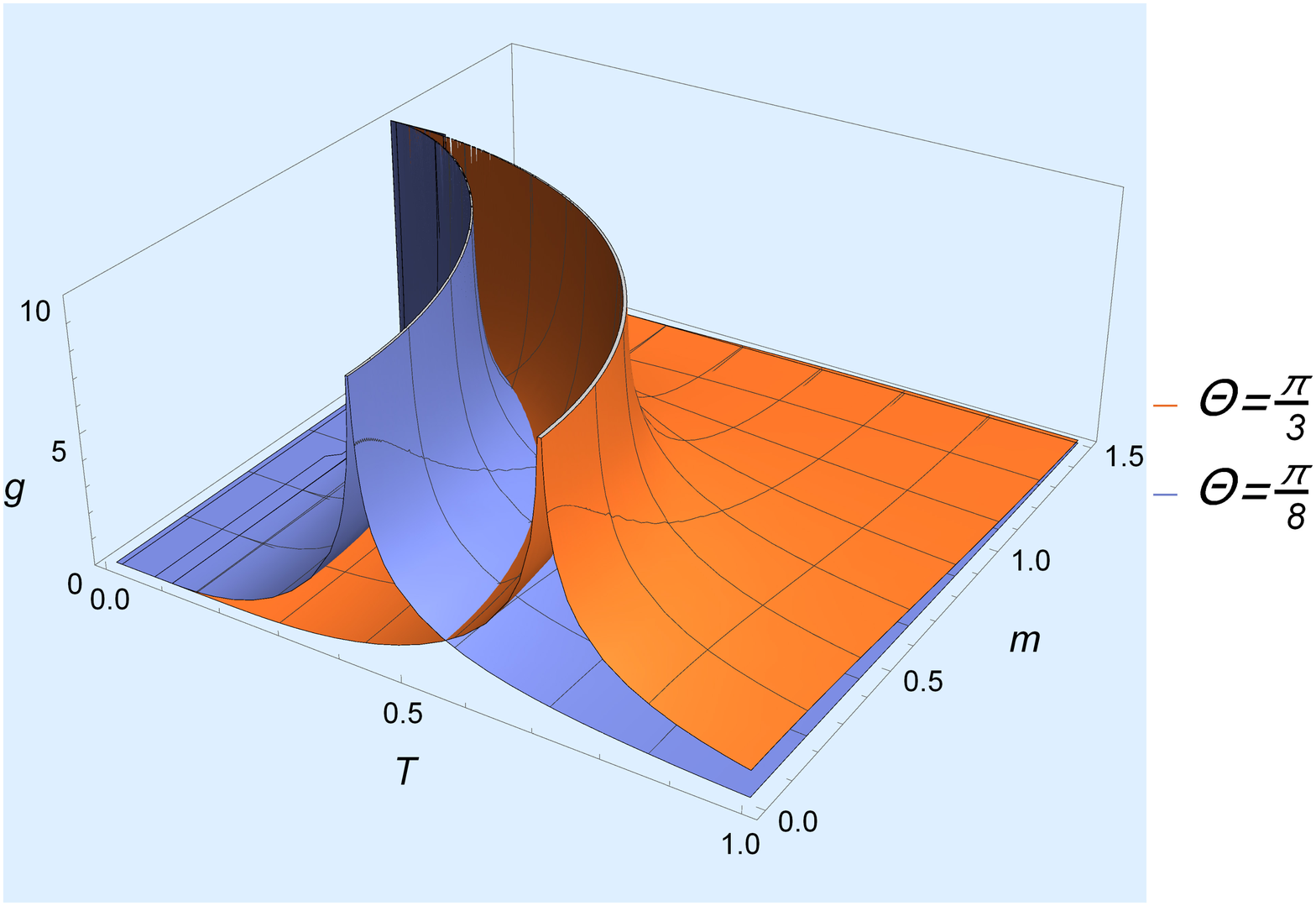}
\includegraphics[width=3.2in]{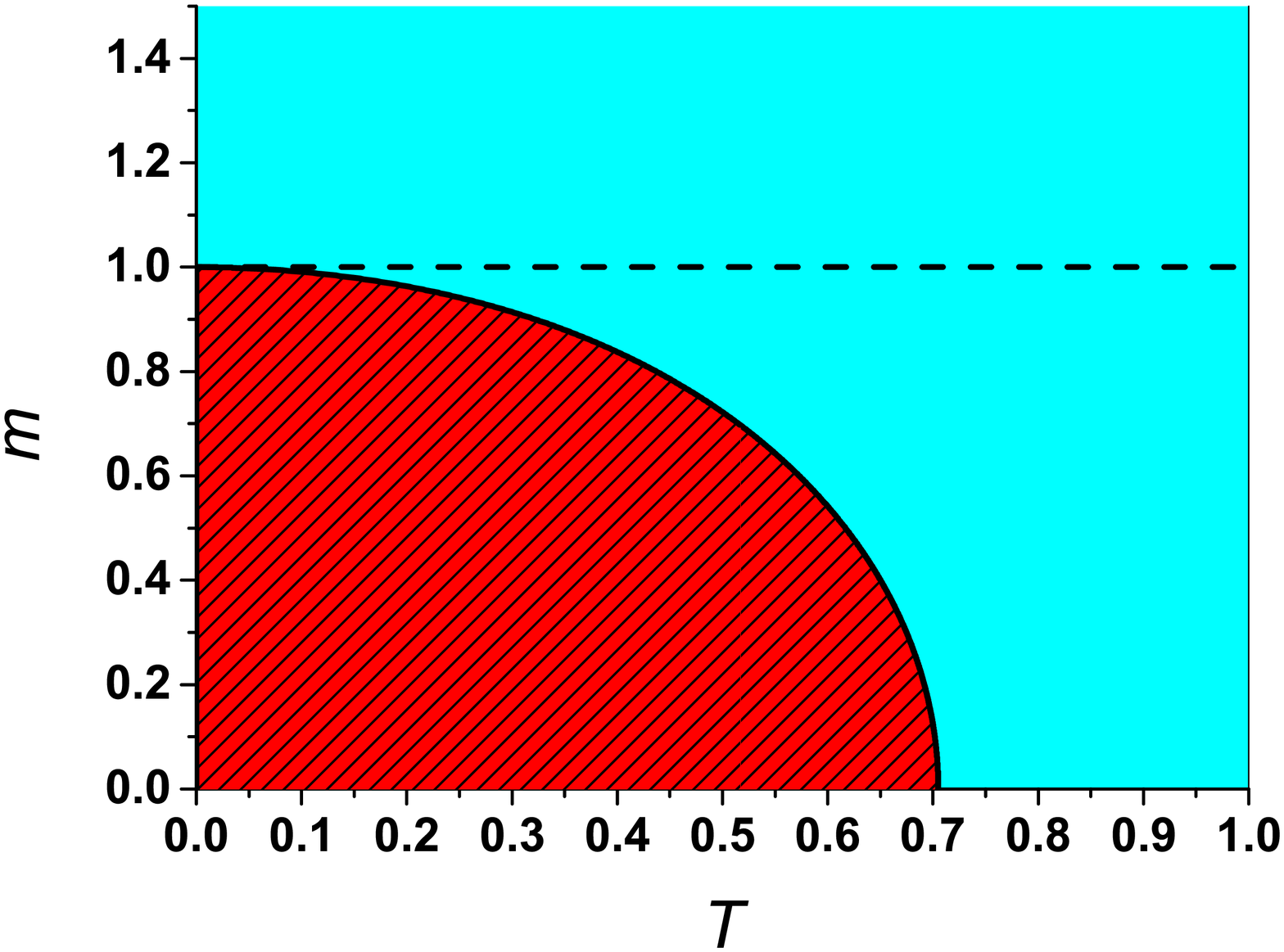}
\caption{(Color online) (Top panel) The geometrical generating function~\eqref{eq:fU} versus $T$ and $m$ for the Creutz ladder going through an Uhlmann process, showing diverging peaks at the finite-temperature TQPTs. The surface with peaks on the right (left) corresponds to $\Theta=\frac{\pi}{3}$ and $\frac{\pi}{8}$, respectively. (Bottom panel) The Uhlmann phase as a function of $T$ and $m$ for $\Theta=\frac{\pi}{3}$. Here $\theta_U=\pi$ in the red shaded (lower) region and $\theta_U=0$ in the blue (upper) region. The dashed line at $m=1.0$ indicates where the winding number $\omega_1$ jumps from $1$ to $0$. }
\label{Fig1}
\end{figure}

It is important to address how the finite-temperature TQPT approaches zero temperature and its relation with the zero-temperature TQPT.
As $T\rightarrow 0$, Eq.~(\ref{GU}) indicates that the Loschmidt amplitude and the Uhlmann phase take the forms
\begin{align}\label{UGT0}
\mathcal{G}^U_\rho(T\rightarrow0)&=\cos \left(\pi \omega_{1}\right),\notag\\
\theta_U(T\rightarrow 0)&=\arg\left[\cos \left(\pi \omega_{1}\right)\right].
\end{align}
It is clear that no matter $\omega_1$ takes the value $0$ or $1$, $\mathcal{G}^U_\rho$ does not vanish. Therefore, the finite-temperature TQPT does not cover the $T=0$ point. However,
$\theta_U=0$ when $\omega_1=0$ and $\theta_U=\pi$ when $\omega_1= 1$, showing that the $T\rightarrow 0$ Uhlmann phase changes value according to the zero-temperature TQPT.
Although the Uhlmann phase approaches the Berry phase as $T\rightarrow 0$, the Uhlmann bundle requires full-rank density matrices and does not cover the pure states at $T=0$~\cite{Asorey19}. Hence, the critical line of the finite-temperature TQPT terminates at a zero-temperature TQPT, but the two types of TQPTs are not compatible as they merge.

To illustrate the two types of TQPTs explained above, we present an explicit example by analyzing the periodic Creutz ladder~\cite{Creutz99,ViyuelaPRL14}. Here we will elucidate the physics from the point of view of the Loschmidt amplitude. The system is a two-legged ladder with cross stitches, and the Hamiltonian is
 \begin{align}\label{CLH}
H_\text{CL}&=\sum_{i=1}^L\big[K(\me^{-\mi\Theta}a^\dagger_{i+1}a_i+\me^{\mi\Theta}b^\dagger_{i+1}b_i+b^\dagger_{i+1}a_i+a^\dagger_{i+1}b_i)\notag\\
&+Ma^\dagger b_i+\text{H.c.}\big],
\end{align}
where $a_i$ and $b_i$ respectively denote the spinless fermionic annihilation operators on
the $i$-th site of the upper and lower chains, $\Theta$ $\in [-\frac{\pi}{2},\frac{\pi}{2}]$ is the magnetic flux, $K$ ($>0$) is the hoping parameter along the horizontal and diagonal links, and $M$ ($>0$) is the hoping parameter along the vertical links. Introducing the parameter $m=\frac{M}{2K}$, $H_\text{CL}$ can be cast into the form (\ref{Hk2}) with
 \begin{align}
\hat{\mathbf{n}}_k&=\frac{2}{\Delta_k}(m+\cos k,0,\sin\Theta\sin k)^T,\label{nk}\\
\Delta_k&=\sqrt{(m+\cos k)^2+\sin^2\Theta\sin^2 k}\label{dk}
\end{align} in units of $2K=1$~\cite{ViyuelaPRL14}. Here we require that $\Theta\neq 0$ to ensure that $\hat{\mathbf{n}}_k$ has at least two nonzero components. We mention that quench dynamics of the Creutz ladder from a zero-temperature initial state has been analyzed in Ref.~\cite{JafariPRB19}, for example.

To analyze the finite-temperature TQPT, we will show that the Uhlmann phase jumps at the zeros of the Loschmidt amplitude. The jump of the Uhlmann phase agrees with that in Ref.~\cite{ViyuelaPRL14}. The zero of the Loschmidt amplitude leads to nonanalytic behavior of the dynamic free energy density. Before presenting the results, we point out some subtleties of the Uhlmann process. The Loschmidt amplitude is obtained by assuming the system is in equilibrium as indicated by Eq.~(\ref{rhok}), but it is known that the Uhlmann process is incompatible with the dynamical process governed by the Hamiltonian~\cite{ourPRB20}. Therefore, the system needs to be in contact with a reservoir to keep it in equilibrium during a Uhlmann process. Moreover, the thermodynamic free energy is well defined for a thermally equilibrium state and does not exhibit nonanalytic behavior when the Loschmidt amplitude crosses a zero in the Uhlmann process. To characterize the TQPTs signified by the Loschmidt-amplitude zeros, we follow the definition of the thermodynamic free energy in the quench process and define its analogue, the geometrical generating function, as
\begin{equation}\label{eq:fU}
g=-\lim_{L\rightarrow \infty}\frac{1}{L}\ln|\mathcal{G}^U_\rho(T)|^2.
\end{equation}
Here the temperature is introduced as the imaginary time. The analogy allows one to see that, as the thermodynamic free energy exhibits nonanalytic behavior at a conventional phase transition, the geometrical generating function exhibits nonanalytic behavior at a TQPT when the Loschmidt amplitude vanishes.

We show the numerical results of the Creutz ladder in Figure.~\ref{Fig1}. The upper panel shows the geometric generating function $g$ as a function of temperature and $m$ according to Eq.~(\ref{eq:fU}) for selected values of $\Theta=\frac{\pi}{3},\frac{\pi}{8}$, respectively.
One can see that $g$ exhibits nonanalytic behavior as the temperature crosses $T^*$, justifying the occurrence of a finite-temperature TQPT.
Across the critical temperature, the value of the Uhlmann phase jumps from $\pi$ to $0$, as shown in the phase diagram in the lower panel. The value $\theta_U=\pi$ in the red shaded area indicates the topologically nontrivial regime, where the amplitude of the purification exhibits the topology of a Mobius strip after a cycle. The value  $\theta_U=0$ in the green area indicates the topologically trivial regime and the topology is a cylinder.

An interesting feature in the upper panel of Fig.~\ref{Fig1} is the merge of the two critical lines as $T\rightarrow 0$. One can infer the location of the merging point from the middle panel as $m=1.0$. However, $m=1.0$ is the critical point of the zero-temperature TQPT, where the value of the winding number (\ref{wne}) of the Hamiltonian mapping changes.
Therefore, $m=1.0$ corresponds to the TQPT of the ground state determined by the homotopy group of the Hamiltonian mapping. Importantly, the energy spectrum becomes gapless if $m=1.0$, according to Eq.~(\ref{dk}).
The ground-state TQPT at $m=1.0$ only concerns the band structure and applies to the system in a pure state, or equivalently at zero temperature. However, the analysis of Eq.~\eqref{UGT0} shows that the Loschmidt amplitude does not vanish as $T\rightarrow 0$. Therefore, the zero-temperature TQPT of the ground state is different in nature from the finite-temperature TQPT of mixed states indicated by the zeros of the Loschmidt amplitude.  Moreover, the winding number (\ref{wne}) only concerns the Hamiltonian and does not vary with temperature. Hence, we use a dashed line to denote $m=1.0$, which corresponds to a TQPT of the ground state at $T=0$, and  a solid line to denote the TQPT of mixed states, where the Loschmidt amplitude vanishes. At finite temperatures, $\theta_U$ jumps at the TQPT of mixed states, not at the $m=1.0$ line. From the analysis of the Cruetz ladder, we see that the zero-temperature TQPT point is the endpoint of the mixed-state TQPT line, but the two types of TQPTs have different characteristics.

\subsection{Three-level system}
\subsubsection{Quasistatic process}
Our previous example of the two-level system in a quasistatic process has showed that the Loschmidt amplitude can only have zeros at infinite temperature. A natural question is whether the Loschmidt amplitude of a quasistatic process can have zeros at finite temperature. In the following, we present a positive answer by expanding the previous two-level model to a three-level system with the Hamiltonian
\begin{align}\label{H3}
H=R\begin{pmatrix}
\sigma_z & 0 \\ 0 & 1
\end{pmatrix}=R\begin{pmatrix}
1 & 0 & 0 \\
0 & -1 & 0 \\
0 & 0 & 1
\end{pmatrix}.
\end{align}
We set $\omega=\frac{R}{\hbar}$ again, and the Loschmidt amplitude is
\begin{align}\label{H3LA}
&\mathcal{G}_\rho(T,t)=
\text{Tr}(\rho(0)\me^{-\frac{\mi}{\hbar}Ht})\notag\\
&=\frac{(2\me^{-\beta R}+\me^{\beta R})\cos(\omega t)-\mi(2\me^{-\beta R}-\me^{\beta R})\sin(\omega t)}{Z(0)}.
\end{align}
When $2\me^{-\beta R}=\me^{\beta R}$, or equivalently $\beta=\frac{\ln 2}{2R}$, the Loschmidt amplitude possesses zeros at a later time $t^*=\frac{\left(n+\frac{1}{2}\right)\pi}{\omega}$.
Hence, the initial amplitude of the density matrix $W(0)$ is orthogonal to the succeeding amplitude $W(t)$ at times $t=t^*$ and temperature $T_q=\frac{2R}{k_B\ln 2}$. We emphasize again that the system is always in equilibrium and the density matrix does not change during a quasistatic process. 
At time $t^*$ and temperature $T_q$, $W(t)$ becomes orthogonal to $W(0)$ as indicated by the Loschmidt-amplitude zero.

The dynamical phase becomes ill-defined when the Loschmidt amplitude vanishes at $t^{\ast}$ and $T_q$ for the three-level system in a quasistatic process. In fact, when the temperature goes from $\frac{2R}{k_B\ln 2} +0^-$ to $\frac{2R}{k_B\ln 2} +0^+$, $\theta_D$ jumps from $(-1)^{n+1}\frac{\pi}{2}$ to $(-1)^{n}\frac{\pi}{2}$ at $t^*=\frac{\left(n+\frac{1}{2}\right)\pi}{\omega}$. The behavior is different from that of $\theta_D$  of the two-level system in a quasistatic process at infinite temperature discussed previously because one cannot cross the infinite temperature. However, it is similar to the behavior of the Uhlmann phase of the two-level system since the Uhlmann phase jumps at the critical temperature $T^*$ corresponding to a TQPT.

\subsubsection{Quench process}
Next, we ask whether there exist finite-temperature DQPTs in a quench process? Here we emphasize again that the temperature of a quench process refers to the initial state in equilibrium. After the quench starts, the system is out of equilibrium and temperature is no longer well-defined.
By slightly modifying the previous example of the three-level system, we will provide a positive answer. The initial Hamiltonian is assumed to be the same as Eq.~(\ref{H3}), and the initial state is in equilibrium with temperature $T=\frac{1}{k_B\beta}$. Thus, the initial density matrix is given by
\begin{align}\label{eq:3Lrho}
\rho(0)&=\frac{1}{Z(0)}\begin{pmatrix}
        \me^{-\beta R} & 0 & 0\\
        0 & \me^{\beta R} & 0\\
        0 & 0 & \me^{-\beta R}
        \end{pmatrix}
\end{align}
where $Z(0)=\text{Tr}\rho(0)$.
At time $t=0^+$, the Hamiltonian is suddenly quenched and becomes
\begin{align}\label{Hf}
H_\text{f}=\begin{pmatrix}
\mathbf{R}\cdot\vec{\sigma} & 0 \\ 0 & R
\end{pmatrix}=R\begin{pmatrix}
    \cos\theta & \sin\theta\me^{-\mi\phi} & 0\\
    \sin\theta\me^{\mi\phi} & -\cos\theta & 0\\
    0 & 0 & 1
    \end{pmatrix}
\end{align}
where 
$\mathbf{R}=R(\sin\theta\cos\phi,\sin\theta\sin\phi,\cos\theta)^T$. A straightforward calculation shows
\begin{align}
\mathcal{G}_\rho(T,t)
&=\frac{1}{Z}\big[
    \cos(\omega t)\left(
        2\me^{-\beta R}+\me^{\beta R}
    \right)\notag\\&
    +\mi\sin(\omega t)\left(
        (-1-\cos\theta)\me^{-\beta R}+
        \cos\theta\me^{\beta R}
    \right)
\big].
\end{align}
The details can be found in the Appendix.
If $(-1-\cos\theta)\me^{-\beta R}+\cos\theta\me^{\beta R}=0$
, i.e.,
$\beta=\frac{ \ln (1+\text{sec}\theta)}{2R}$, then $\mathcal{G}_\rho(T,t)$ possesses zeros at times $t^*=\frac{(n+\frac12)\pi}{\omega}$ with $n$ being an integer. To ensure the temperature is positive, it is required that $\theta\in[0,\frac{\pi}{2})$. Thus, the DQPTs of the three-level system occur at temperature $T_h=\frac{2R}{k_B \ln (1+\text{sec}\theta)}$. After some algebra, the dynamical phase can be shown to jump from $(-1)^{n+1}\frac{\pi}{2}$ to $(-1)^{n}\frac{\pi}{2}$ at $t^*=\frac{\left(n+\frac{1}{2}\right)\pi}{\omega}$ as the temperature crosses the critical value $T_h$.

\begin{figure}[th]
\centering
\includegraphics[width=3.4in,clip]
{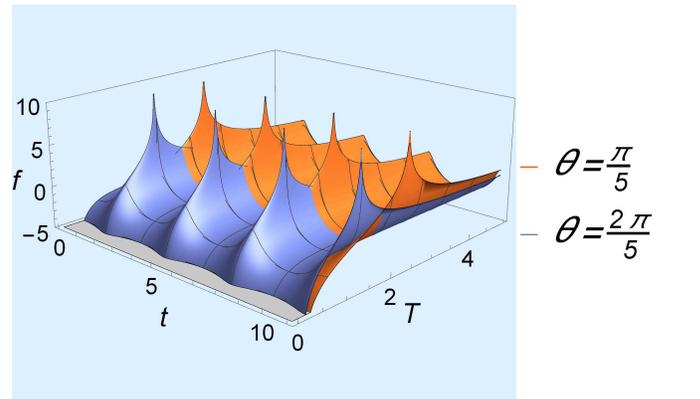}
\caption{(Color online) Dynamical free energy density $f$, defined in Eq.~\eqref{eq:dyfe}, as a function of $t$ and $T$ for the three-level system in a quench process. The diverging peaks show where the finite-temperature DQPTs occur. The brown and blue surfaces correspond to $\theta=\frac{\pi}{5}$ and $\frac{2\pi}{5}$, respectively. }
\label{Fig2}
\end{figure}

To visualize the results, we plot the dynamical free energy density $f$, defined in Eq.~\eqref{eq:dyfe}, of the three-level system in the quench process as a function of time and temperature in Figure.~\ref{Fig2} with two selected values of the parameter $\theta$. One can see that $f$ diverges at a series of values of $t^*$ and finite temperature $T_h$, indicating the occurrence of finite-temperature DQPTs. When the temperature crosses the divergent points of $f$, the value of the dynamical phase jumps by $\pi$, as pointed out previously.

\subsubsection{Uhlmann process}
We analyze a three-level system undergoing an Uhlmann process to compare with the Cruetz ladder. The three-level system has the Hamiltonian given by Eq.~(\ref{Hf}), but the Hamiltonian is time-independent for the Uhlmann process. 
The parameter space is determined by $(\theta,\phi)$, corresponding to a 2D sphere. To evaluate the Loschmidt amplitude according to Eq.~(\ref{GrU}), we consider the case where the system traverses a simple loop in the parameter space described by the circle of latitude with $\theta=\frac{\pi}{2}$. Under this condition, the Uhlmann connection takes a simple form and the path-ordered integral in Eq.~(\ref{GrU}) can be explicitly evaluated. A tedious but straightforward evaluation by using Eq.~(\ref{GrAU}) then gives
\begin{align}
A_{U}=\frac{\mi}{2} \frac{\me^{ -\beta R}+\me^{\beta R}-2}{\me^{ -\beta R}+\me^{\beta R}}
\begin{pmatrix}
\sigma_z & 0\\
0 & 0
\end{pmatrix} \dif\phi.
\end{align}
The system is assumed to start from $\phi=0$. The initial density matrix is then given by 
\begin{align}
\rho(0)=\frac{\me^{-\beta H(0)}}{Z(0)}=\frac{1}{Z(0)}\exp[-\beta R\begin{pmatrix}
\sigma_x & 0\\
0 & 1
\end{pmatrix}].
\end{align}
Substituting those into Eq.~(\ref{GrU}), the Loschmidt amplitude is given by
\begin{align}\label{GU3}
\mathcal{G}^U_\rho(T)=\frac{1}{Z(0)}\left[-2\cosh(\beta R)\cos\left( \frac{\pi
}
{\cosh(\beta R)}\right)+\me^{-\beta R} \right],
\end{align}
which possesses a zero at $T^*\approx 0.7338\frac{R}{k_B}$. The zero of the Loschmidt amplitude indicates a finite-temperature TQPT of the three-level system.

\begin{figure}[th]
\centering
\includegraphics[width=3.4in,clip]
{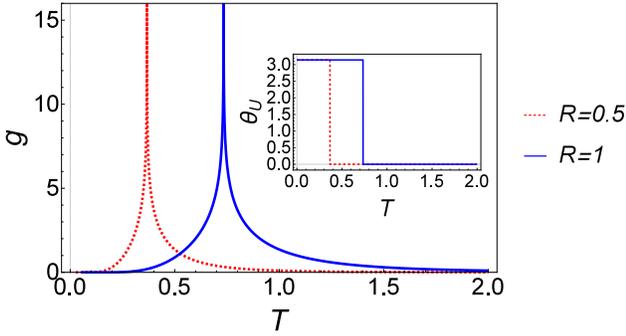}
\caption{(Color online)  The geometrical generating function $g$~\eqref{eq:fU} of the three-level system as a function of $T$. The red dotted and blue solid lines correspond to $R=0.5$ and $1.0$, respectively. The diverging peaks indicate where the finite-temperature TQPTs occur. The inset shows the jump of the Uhlmann phase at the critical temperature.}
\label{Fig3}
\end{figure}

To better understand the result, we plot the geometrical generating function $g$, defined in Eq.~(\ref{eq:fU}), versus temperature in Figure~\ref{Fig3}. One clearly sees that $g$ diverges at $T^*$, indicating the finite-temperature TQPT. Interestingly, the Uhlmann phase, which is the argument of $\mathcal{G}^U_\rho(T)$ as shown in Eq.~(\ref{GPU}), jumps from $\pi$ to 0 when the temperature goes from $T^* +0^-$ to $T^* +0^+$. The inset of Figure~\ref{Fig3} shows the jump of the Uhlmann phase. Similar to the two-band Creutz ladder discussed previously, the three-level system has a topological phase at low temperature with a quantized Uhlmann phase and topologically trivial phase at high temperature. Therefore, the zero of the Loschmidt amplitude provides a general description of finite-temperature TQPTs. In contrast to the periodic peaks of the DQPTs of the three-level system shown in Fig.~\ref{Fig2}, there is only one finite-temperature TQPT of the three-level system with a fixed set of parameters because the Hamiltonian is time-independent through the Uhlmann process.

\section{Representation of purification of two-level systems}\label{RP}
\subsection{Purified states as spinors}
Here we clarify, in a more concrete way, the role of the amplitude of a density matrix in purification by using the two-level system as an example. The Hamiltonian is $H=\mathbf{R}\cdot\vec{\sigma}$ and the corresponding density matrix in thermal equilibrium is given by Eq.~(\ref{rho2l}). We introduce $\Delta=2 R$, which determines the energy gap between the two levels. The density matrix can be rewritten as $\rho=\frac{1}{2}(1-\tanh(\frac{\beta \Delta}{2})\vec{\sigma}\cdot\hat{\mathbf{R}})$. We will use the properties of the Fermi distribution function $n_{\text{f}}(x)=1/(\me^{\beta x}+1)$ to find the square root of $\rho$ with respect to the outer-product. One can show that 
 \begin{align}\label{FD}
&n_{\text{f}}(\Delta)+n_{\text{f}}(-\Delta)=1, \notag\\
& n_{\text{f}}(\Delta)-n_{\text{f}}(-\Delta)=-\tanh\left(\frac{\beta \Delta}{2}\right).
\end{align}
Hence, the density matrix is a linear combination of the two projection operators
\begin{align}\label{PO}
P_\pm=\frac{1}{2}\left(1\pm\vec{\sigma}\cdot\hat{\mathbf{R}}\right)=\frac{1}{2}\left(1\pm\frac{H}{R}\right)
\end{align}
that project into the states $|\pm R\rangle$. Explicitly,
\begin{align}\label{rhoP}
\rho=n_{\text{f}}(\Delta)P_++n_{\text{f}}(-\Delta)P_-.
\end{align}
We remark that $P_\pm$ are indeed projectors since $P^2_+=P^2_-=1$, and are orthogonal to each other since $P_+P_-=P_-P_+=0$.

The two energy levels may be parameterized by
\begin{align}\label{EL2}
|+R\rangle=\left(\begin{array}{c}
\cos\frac{\theta}{2}\\ \sin\frac{\theta}{2}\me^{\mi\phi}
 \end{array}\right),\quad
|-R\rangle=\left(\begin{array}{c}
\sin\frac{\theta}{2}\\ -\cos\frac{\theta}{2}\me^{\mi\phi}
 \end{array}\right).
\end{align}
A straightforward evaluation shows
\begin{align}\label{soPO}
&P_+=\frac{1}{2}
\begin{pmatrix}
1+\cos\theta & \sin\theta\me^{-\mi\phi}\\ \sin\theta\me^{\mi\phi} & 1-\cos\theta
\end{pmatrix}
=|+R\rangle\langle +R|
,\notag\\
&P_-=\frac{1}{2}
\begin{pmatrix}
1-\cos\theta & -\sin\theta\me^{-\mi\phi}\\ -\sin\theta\me^{\mi\phi} & 1+\cos\theta
\end{pmatrix}
=|-R\rangle\langle -R|.
\end{align}
Hence, $P_\pm$ are actually the density matrices of the pure states $|\pm R\rangle$.
On the other hand, $|\pm R\rangle$ also represents $P_\pm$ in the Bloch-sphere representation of the density matrices of two-level systems~\cite{CQI_book}.
Eq.~(\ref{rhoP}) indicates that the eigenvalues of $\rho$ are $n_\text{f}(\pm \Delta)$. According to Eq.~(\ref{w2}),
the purified state $|W\rangle$ of $\rho$ can be written as
\begin{align}\label{PO2}
|W\rangle=\frac{1}{\sqrt{\me^{\beta \Delta}+1}}|W_+\rangle+\frac{1}{\sqrt{\me^{-\beta \Delta}+1}}|W_-\rangle,
\end{align}
where
\begin{align}\label{POrho}
|W_+\rangle&=|+R\rangle\otimes |+R\rangle
=\frac{1}{2}\begin{pmatrix}
1+\cos\theta\\ \sin\theta\me^{\mi\phi} \\ \sin\theta\me^{\mi\phi} \\
(1-\cos\theta)\me^{2\mi\phi}
\end{pmatrix},\notag\\
|W_-\rangle&=|-R\rangle\otimes |-R\rangle
=\frac{1}{2}\begin{pmatrix}
1-\cos\theta\\ -\sin\theta\me^{\mi\phi} \\ -\sin\theta\me^{\mi\phi} \\
(1+\cos\theta)\me^{2\mi\phi}
\end{pmatrix}.
\end{align}

To clarify the physical role of $|W\rangle$, we extend the unit vector $\hat{\mathbf{R}}$ to a Minkowski 4-vector $R^\mu=(1,\hat{\mathbf{R}})$. The introduction of a relativistic structure actually helps us find the square roots with respect to the outer product, which will be shown shortly.
The Minkowski metric tensor is taken as $\eta_{\mu\nu}=\eta^{\mu\nu}=\text{diag}(1,-1,-1,-1)$. Hence, the norm of $R^\mu$ is zero since $R^\mu R_\mu=0$.
To further simplify the notations, we define $\sigma^\mu=(1,\vec{\sigma})$ and $\bar{\sigma}^\mu=(1,-\vec{\sigma})$. Then,  $P_+=\frac{1}{2}R^\mu\bar{\sigma}_\mu$ and $P_-=\frac{1}{2}R^\mu\sigma_\mu$. In other words, $P_\pm$ are two matrix representations of the zero-norm 4-vector $R^\mu$ and satisfy
 \begin{align}\label{Ppm}
\det P_+=\det P_-=\frac{1}{4}R^\mu R_\mu=\frac{1}{4}\left(1-\hat{\mathbf{R}}^2\right)=0.
\end{align}
In the $2\times 2$ matrix representation of 4-vectors, only those with zero norm allows a square-root decomposition~\cite{HBY_book,Spinor}.
Since $P_\pm$ can be ``factorized'' as $P_\pm=|\pm R\rangle\langle \pm R|$, mathematically $|\pm R\rangle$ can be thought of as the ``square roots'' of the zero-determinant matrices $P_\pm$~\cite{HBY_book,Spinor}, respectively. We remark that the square root discussed here is based on the operation of the outer product of the vectors. Symbolically, this implies that $|\pm R\rangle$ are the two ``square roots'' of $R^\mu$ in the sense that $P_\pm$ form a matrix representation of $R^\mu$. Moreover,
since $|W_\pm\rangle\langle W_\pm|=P_\pm\otimes P_\pm$ is a reducible matrix representation of $R^\mu$,
$|W_\pm\rangle$ are also the two square roots of $R^\mu$ in higher dimensions by the same reasoning.

Finally, the role of $|W\rangle$, given in Eq.~(\ref{PO2}), needs a more careful investigation.
Symbolically, the square root of a vector, or more precisely, ``half of a vector''~\cite{Spinor} with respect to the direct product, is recognized as a spinor. Therefore, $|\pm R\rangle$ and $|W_\pm\rangle$ are both spinors. $|W_\pm\rangle$ are four-component spinors that are reducible since $|W_\pm\rangle=|\pm R\rangle\otimes |\pm R\rangle$. On the other hand, $|\pm R\rangle$ are two-component spinors that are irreducible.
We remark that the reason a zero-norm 4-vector has two square roots in the 2D matrix representation is because of the mathematical proposition SO(1,3)=SU(2)$\times$SU(2)/$Z_2$~\cite{HBY_book}.

We now come back to the density matrix of an arbitrary mixed state given by Eq.~(\ref{rhoP}) and analyze the role of its square root. Since $\det\rho=\frac{1}{4}\text{sech}^2\left(\frac{\beta \Delta}{2}\right)>0$, $\rho$ can not be the 2D matrix representation of any zero-norm 4-vector. This is consistent with the fact that $\rho$ of non-pure states has no two-component square roots, i.e., $\rho\neq |\psi\rangle\langle \psi|$. However, if we introduce the 4-vector $\tilde{R}^\mu=(1,\tanh(\frac{\beta\Delta}{2}))$ with a nonzero norm, then $\rho$ is the 2D matrix representation of $\tilde{R}^\mu$ because $\rho=\frac{1}{2}\tilde{R}^\mu\bar{\sigma}_\mu$. Thus, $\rho$ (or $\tilde{R}^\mu$) only has four-component square roots manifested by $\rho=\text{Tr}_2(|W\rangle\langle W|)$. Here $|W\rangle$ is a four-component spinor given by Eq.~(\ref{PO2}), which cannot be
reduced to a two-component spinor.

To summarize, if $\rho$ denotes the density matrix of a pure state, it is a matrix representation of a 4-vector with zero norm. Hence, its purification is a four-component spinor, which can be further reduced to a two-component spinor. This is consistent with the fact that its purification is a separable state.
On the other hand, if $\rho$ is the density matrix of a non-pure mixed state, it is a matrix representation of a 4-vector with nonzero norm. Thus, its purification is a four-component irreducible spinor. This is consistent with the observation that its purification results in an entangled state.

It is possible that purification of the density matrix of systems with more than two energy levels may be represented by some kind of spinors of higher dimensions. However,
a rigorous proof of the proposition is beyond the scope of the paper and awaits future research.

\subsection{Quasistatic processes as rotations}
We give an example of how the spinor representation of a two-level system helps us visualize the evolution by showing that the purified state rotates as the two-level system undergoes a quasistatic process. The purification evolves during a quasistatic process according to Eq.~\eqref{Wt},
and the time-evolution operator can be expressed as
\begin{align}
\me^{-\mi Ht}\otimes 1=\me^{-\mi R_i\sigma_i t}\otimes 1=\me^{-\mi R_it(\sigma_i\otimes 1) }=\me^{-\mi R_it\Gamma_i }.
\end{align}
Here $\Gamma_i\equiv\sigma_i\otimes 1$ are the gamma matrices (with $\sigma_i\otimes 1$ being a representation of the gamma matrices). Hence, the quasistatic process introduces a rotational operator acting on the four-component spinor of the purified state. Physically, the purification of the same density matrix is rotated like a spinor by the operator $\me^{-\mi Ht}$ during a quasistatic process. Note the density matrix changes during a quench or Uhlmann process, so those processes do not fit the situation considered here.

If the initial purified state $|W(0)\rangle$ is given by Eq.~(\ref{PO2}) and we apply Eq.~(\ref{H2l}), the succeeding purified state at time $t$ is given by
\begin{align}
|W(t)\rangle=\frac{\me^{-\mi\omega t}}{\sqrt{\me^{\beta \Delta}+1}}|W_+\rangle+\frac{\me^{\mi\omega t}}{\sqrt{\me^{-\beta \Delta}+1}}|W_-\rangle.
\end{align}
It is still a purification of the initial density matrix, i.e.,
\begin{align}
\text{Tr}_2(|W(t)\rangle\langle W(t)|)=\text{Tr}_2(|W(0)\rangle\langle W(0)|)=\rho(0)
\end{align}
since $\text{Tr}_2(|W_+\rangle\langle W_-|)=\text{Tr}_2(|W_-\rangle\langle W_+|)=0$. The Loschmidt amplitude of the quasistatic process is then given by
\begin{align}
\mathcal{G}_\rho(T,t)=\langle W(0)|W(t)\rangle=\frac{\me^{-\mi\omega t}}{\me^{\beta \Delta}+1}+\frac{\me^{\mi\omega t}}{\me^{-\beta \Delta}+1}.
\end{align}
One can verify that only at infinite temperature, $\beta\rightarrow 0$, $\mathcal{G}_\rho(T,t)=\cos(\omega t)$ and has zeros at  $t^*=\frac{(n+\frac12)\pi}{\omega}$ with integer $n$. It implies that $|W(t)\rangle$ has been rotated to an orthogonal state of $|W(0)\rangle$ at $t^*$. This description complements our previous discussions, and the spinor representation of purification of two-level systems offers a picture for visualizing the physics.

\section{Conclusion}\label{Sec5}
The ubiquity of the zeros of the Loschmidt amplitude of mixed quantum states undergoing different physical processes has been demonstrated via the concept of purification, the overlap of the purified states, and the dynamical or geometrical phase of the corresponding process. For a quasistatic process, the purified state becomes orthogonal to the initial one at the Loschmidt-amplitude zero, and the dynamical phase jumps despite the fact that both states purify the same density matrix. The well-defined thermodynamic free energy in a quasistatic process, however, rules out the association of the Loschmidt-amplitude zero with a phase transition.

Nevertheless, the DQPT of a quench process occurs at the zero of the Loschmidt amplitude due to the nonanalytic behavior in the dynamical free energy, accompanied by a jump of the dynamical phase. While DQPTs at infinite temperature have been proposed in two-level systems, we demonstrate a finite-temperature DQPT of a three-level system. Finally, for an Uhlmann process, there may be two types of TQPTs, one at zero temperature associated with the topology of the Hamiltonian mapping and the other at finite temperature associated with the zeros of the Loschmidt amplitude. For the Cruetz-ladder model, the critical line of the finite-temperature TQPT terminates at the critical point of the zero-temperature TQPT. The Uhlmann process of the three-level system also exhibits a finite-temperature TQPT. Moreover, the Uhlmann phase jumps when the system crosses the Loschmidt-amplitude zeros.

The phase transitions in thermodynamics occur because of nonanalytic behavior of the thermodynamic free energy. The generalizations of the concept to nonanalytic behavior of the overlap between quantum states offer a framework for connecting the seemingly different dynamical and geometrical processes from zero to infinite temperatures. The spinor-representation of the purified states of two-level systems further elucidates the structure behind purification in different physical processes. Applications of the framework and concept to more complicated systems may bring forth more examples that bridge the descriptions of quantum phenomena in and out of equilibrium.

\begin{acknowledgments}
H. G. was supported by the National Natural Science Foundation of China (Grant No. 11674051).
\end{acknowledgments}

\appendix
\section{Details of the Loschmidt amplitude}
A derivation of Eq.~(\ref{GU}) is outlined here. The Loschmidt amplitude in an Uhlmann process is given by
\begin{align}\label{tmp1}
\mathcal{G}^U_\rho(T)&=\langle W_{k(0)}|W_{k(1)}=\textrm{Tr}[W_{k(0)}^\dagger W_{k(1)}]\notag\\
&=\textrm{Tr}[\rho_{k(0)}\mathcal{P}\me^{-\oint A_U}],
\end{align}
where $\{k(t)\}_{t=0}^1$ denotes a closed path in the Brillouin
zone, and $A_U$ is given by Eq.~(\ref{GrAU}). To find the expression of $\mathcal{G}^U$ for the general two-level system, we plug Eq.~(\ref{Hk2}) into Eq.~(\ref{rhok}) and get 
\begin{widetext}
\begin{align}
\rho_k=
\begin{pmatrix}n_\text{f}(\Delta_k)+\mathrm{coth}\frac{\beta\Delta_k}{2}\sin^2\frac{\theta_k}{2} &
\frac{1}{2}\mathrm{coth}\frac{\beta\Delta_k}{2}\me^{-\mi\phi_k}\sin\theta_k\\
\frac{1}{2}\mathrm{coth}\frac{\beta\Delta_k}{2}\me^{\mi\phi_k}\sin\theta_k&
n_\text{f}(\Delta_k)+\mathrm{coth}\frac{\beta\Delta_k}{2}\cos^2\frac{\theta_k}{2}\end{pmatrix},
\end{align}
\end{widetext}
where $\hat{n}_k=(\sin\theta_k\cos\phi_k,\sin\theta_k\sin\phi_k,\cos\theta_k)^T$. To simplify the calculation, the closed path $\{k(t)\}_{t=0}^1$ is chosen so that $\hat{n}_k$ stays on a plane. A simple choice is either $\phi_k=0$ or $\theta_k=\frac{\pi}{2}$. Here the former is adopted.
By using Eq.~(\ref{GrAU}), a lengthy but straightforward calculation leads to 
\begin{align}\label{Pua2}
\mathcal{G}^U_\rho(T)=\cos\big[\frac{1}{2}\oint\mathrm{d}\theta_k-\frac{1}{2}\oint\mathrm{sech}\frac{\beta\Delta_k}{2}\mathrm{d}\theta_k\big].
\end{align}
Note that
\begin{align}\label{t8}
\omega_1=\frac{1}{2\pi}\oint\mathrm{d}\theta_k=\frac{1}{2 \pi} \oint\left(\frac{\partial_{k} n_{k}^{i}}{n_{k}^{j}}\right) \dif k
\end{align}
is the winding number shown in Eq.~(\ref{wne}). Then $\sin(\pi\omega_1)=0$ because of the quantization of the winding number. Thus, Eq.~(\ref{Pua2}) finally leads to Eq.~\eqref{GU}

For the three-level system undergoing a quench process, the quenched Hamiltonian given by Eq.~(\ref{Hf}) can be diagonalized as
\begin{align}
    H_\text{f}=S\begin{pmatrix}
        R & 0 & 0\\
        0 & -R & 0\\
        0 & 0 & R
        \end{pmatrix}
        S^{-1},
    \end{align}
    where
    \begin{align}
    S&=\begin{pmatrix}
        \cos\frac{\theta}{2} & \sin\frac{\theta}{2} & 0\\
        \sin\frac{\theta}{2}\me^{\mi\phi} & -\cos\frac{\theta}{2}\me^{\mi\phi} & 0\\
        0 & 0 & 1
        \end{pmatrix},\notag\\
    S^{-1}&=\begin{pmatrix}
        \cos\frac{\theta}{2} & \sin\frac{\theta}{2}\me^{-\mi\phi} & 0\\
        \sin\frac{\theta}{2} & -\cos\frac{\theta}{2}\me^{-\mi\phi} & 0\\
        0 & 0 & 1
        \end{pmatrix},
\end{align}
and the initial density matrix has been shown in Eq.~\eqref{eq:3Lrho}.
From those expressions, the Loschmidt amplitude is given by
\begin{widetext}
\begin{align}
\mathcal{G}_\rho(T,t)
&=\frac{1}{Z(0)}\text{Tr}\left[\rho(0) S\begin{pmatrix}
    \me^{-\frac\mi\hbar Rt} & 0 & 0\\
    0 & \me^{\frac \mi\hbar Rt} & 0\\
    0 & 0 & \me^{-\frac\mi\hbar Rt}
    \end{pmatrix}S^{-1}
    \right]\notag\\
&=\frac{1}{Z}\left[\left(\cos^{2}\frac{\theta}{2}\me^{-\beta R}+
\sin^{2}\frac{\theta}{2}\me^{\beta R}\right)
\me^{-\mi\omega t} +
\left(\sin^{2}\frac{\theta}{2}\me^{-\beta R}
+\cos^{2}\frac{\theta}{2}\me^{\beta R}\right)\me^{\mi\omega t}+
\me^{-\beta R}
\me^{-\mi\omega t}\right]\notag\\
&=\frac{1}{Z}\left[
    \cos(\omega t)\left(
        2\me^{-\beta R}+\me^{\beta R}
    \right)
    +\mi\sin(\omega t)\left(
        (-1-\cos\theta)\me^{-\beta R}+
        \cos\theta\me^{\beta R}
    \right)
\right].
\end{align}
\end{widetext}

\bibliographystyle{apsrev}

\end{document}